\documentstyle[12pt,aasms4]{article}

\vspace*{-0.3cm}
\slugcomment{submitted to the Ap.J.}

\lefthead{Della Valle et al.}
\righthead{SNIa in Radiogalaxies}

\newcommand{\ie}{\emph{i.e.}\ }
\newcommand{\etal}{\emph{et al.}\ }
\newcommand{\eg}{\emph{e.g.}\ }

\newcommand{\msun}{M$_\odot$}

\newcommand{\lsim}{{\, \lower2truept\hbox{
${< \atop\hbox{\raise4truept\hbox{$\sim$}}}$}\,}}
\newcommand{\gsim}{{\, \lower2truept\hbox{
${> \atop\hbox{\raise4truept\hbox{$\sim$}}}$}\,}}

\begin{document}

\title{Why Are Radio-Galaxies Prolific Producers of Type 
Ia Supernovae?$^\dag$}
\renewcommand{\thefootnote}{\fnsymbol{footnote}}\footnotetext[2]{{\bf
Version 6 of \today}} \renewcommand{\thefootnote}{\arabic{footnote}}

\author{Massimo Della Valle}
\affil{Istituto Nazionale di Astrofisica, Osservatorio Astrofisico di
Arcetri,\\ Largo Enrico Fermi 5,
I-50125 Florence, Italy}

\author{Nino Panagia\altaffilmark{1} }
\affil{Space Telescope Science Institute, 3700 San Martin Drive,
Baltimore, MD~21218}

\author{Paolo Padovani}
\affil{European Southern Observatory, Karl-Schwarzschild-Strasse 2,
Garching bei M\"unchen, Germany}

\author{Enrico Cappellaro} \affil{Istituto Nazionale di Astrofisica,
Osservatorio Astronomico di Capodimonte\\ Salita Moiariello,
16, I-80131, Napoli, Italy}

\author{Filippo Mannucci}
\affil{CNR-IRA\\ Largo Enrico Fermi 5,
I-50125 Florence, Italy}

\author{Massimo Turatto}
\affil{Istituto Nazionale di Astrofisica, Osservatorio Astronomico di
Padova,\\ Vicolo Osservatorio 5, Padova, Italy}

\altaffiltext{1}{On assignment from the Space Telescope Operations
Division, Research and Scientific Support Department of ESA}

\begin{abstract}

An analysis of SNIa events in early type galaxies from the Cappellaro
\etal~(1999) database provides conclusive evidence that the rate of
type Ia Supernovae (SNe) in radio-loud galaxies is about 4 times
higher than the rate measured in radio-quiet galaxies, \ie~
SNIa-rate$(radio-loud~galaxies)=0.43^{+0.19}_{-0.14}h^2_{75}$ SNu as
compared to
SNIa-rate$(radio-quiet~galaxies)=0.11^{+0.06}_{-0.03}h^2_{75}$ SNu.
The actual value of the enhancement is likely to be in the range
$\sim 2-7$ (P$\sim 10^{-4}$). This finding puts on robust empirical
grounds the results obtained by Della Valle \& Panagia (2003) on the
basis of a smaller sample of SNe.

We analyse the possible causes of this result and conclude that
the enhancement of SNIa explosion rate in radio-loud galaxies has
the same origin as their being strong radio sources, but
there is no causality link between the two phenomena.  We argue that
repeated episodes of interaction and/or mergers of early type
galaxies with dwarf companions, on times-scale of about 1~Gyr, are
responsible for inducing both strong radio activity observed in
$\sim$14\% of early type galaxies and to supply an adequate number of
SNIa progenitors to the stellar population of ellipticals.
\end{abstract}


\keywords{(stars:) stars: supernovae -- galaxies: radio}

\section{Introduction}
\bigskip

The Supernova rate is a key quantity for astrophysics. Its knowledge is
essential to constrain the mechanisms of galaxy formation and for
understanding galaxy evolution. In particular, the chemical evolution
of systems that at the present epoch have substantially reduced or
totally exhausted star formation activity, such as elliptical
galaxies, may dramatically depend on the rate of type Ia events
(e.g. Renzini et al. 1993, Maoz \& Gal-Yam 2004). Thus, any mechanism
capable of producing a substantial enhancement of SNIa rate in this
type of galaxies should be carefully investigated. Moreover, 
studies of the connection between the frequency
of occurrence of type Ia SNe and the properties of the environment
where the SNe-Ia originate (e.g. Della Valle \& Livio 1994; Mannucci
et al. 2005) can effectively constraint the still controversial nature
of their progenitors (e.g. Hamuy et al. 2003, Livio and Riess 2003).

In recent years three works have dealt with this topic, namely Capetti (2002),
Livio, Riess and Sparks (2002) and Della Valle and Panagia (2003). The
former two papers claim the existence of a {\sl causality} link
between the position of 6 type Ia SNe (Capetti) and a dozen of nova
stars (Livio et al.) with the jets of the respective parent
radio-galaxies. In this scenario it is the jet itself that modifies
the physical conditions of its environment by increasing the accretion
rate from the interstellar medium onto the white dwarf progenitors,
and thus it is the {\it direct cause} of a local enhancement of Nova
and type Ia SN explosions.

Della Valle \& Panagia, after analysing the Evans, van den Bergh \&
Clure (1989) database (hereafter EVC89), provided strong
evidence for the existence of an enhancement by a factor 4--15 of the
SN-Ia rate in radio-loud galaxies with respect to radio-quiet
ones. They argued that the mechanism which boosts the SN production is
essentially independent of the position of the Supernovae inside the
parent galaxy, rather having the same common origin as the parent
galaxies being strong radio sources, {\sl without causality} link
between the two phenomena. They suggested that repeated episodes of
interaction and/or mergers of early type galaxies with dwarf
companions are responsible for inducing both strong radio activity in
$\sim$14\% of early type galaxies and the $\sim1$~Gyr old stellar
population needed to supply an adequate number SNIa progenitors. 
However, their results were based on a relatively small sample and deserve
further investigation. 


In this paper we analyse an approximately five times larger sample of SNe
occurred in early type galaxies, as reported by Cappellaro, Evans \&
Turatto (1999). This sample was obtained by merging the Cappellaro et
al. (1993, hereafter C93) sample with EVC89. In \S~2 we present
the dataset, in \S~3 we first analyse the C93 sample, with the aim of
providing an independent check of the results obtained by Della Valle
\& Panagia (2003) (based on EVC89), then we compute the SN rate in
radio-loud and radio-quiet galaxies on the whole sample of SNe
provided by Cappellaro et al. (1999). In \S~4 we discuss the
results and their implications on the nature of SNIa progenitors.

Tab.~1 gives the comparison between the EVC89 and Cappellaro et
al. (1999) samples. Data for the C93 sample can be obtained by subtracting the EVC89 data
from Cappellaro et al. (1999).

In the following, we will denote as {\sl radio-loud} sources, all
galaxies which have radio luminosities at 1.4GHz higher than $10^{29}$
erg~s$^{-1}$~Hz$^{-1}$ (e.g. Sadler, Jenkins, \& Kotanyi 1989, Ledlow
\& Owen 1996).  The threshold value of $\sim 10^{29}$ erg $s^{-1}$
Hz$^{-1}$ is the faint limit of the luminosity function for the
radio-galaxies (see also Urry \& Padovani 1995), and, therefore, 
represents the `natural' threshold to identify ``radio-loud''
galaxies. We also define {\sl radio-faint} the galaxies characterized
by $4\times 10^{27} \leq P \leq 10^{29}$
erg $s^{-1}$ Hz$^{-1}$, i.e. galaxies whose radio emission is still
not dominated by thermal processes (Phillips et al. 1986; and Sadler
et al. 1989) and finally {\sl radio-quiet} all galaxies with radio
luminosities below $P=4\times 10^{27}$ erg s$^{-1}$ Hz$^{-1}$.
\bigskip

\begin{table}
\begin{center}
\caption{SN samples}
\label{sample}
\begin{tabular}{lrccrcr}
\hline
              &     & Evans          &     &  & Cappellaro   &      \\
galaxies      & $N$ & Control Time & SNe & $N$ & Control Time & SNe \\
\hline
radio-quiet   & 178 & 2270         & 0   & 1738& ~7127         & 7.5  \\
radio-faint$^1$& -- &  --          & --  & 212 & ~1770         & 4.0\\
radio-loud    &  19 & ~847         & 4   & 267 & ~2199         & 9.5  \\
total         & 197 & 3117         & 4   & 2217& 11096         & 21  \\
\hline
\end{tabular}

$^1${\small {Due to the scanty statistic the {\sl
radio-faint} subclass was not adopted in our previous paper (Della Valle
\& Panagia 2003). Radio-faint galaxies
were included in the radio-quiet subclass.}}
\end{center}
\end{table}

\section{Statistics of SNIa explosions in early type galaxies}

\bigskip

The database compiled by Cappellaro et al. (1999) has the advantage of
being defined `a priori', i.e. the monitored galaxies were selected
before the SN explosions were discovered. As a consequence our
statistical analysis will not suffer of any kind of `a posteriori'
selection criteria, which actually may affect the statistical analysis
carried out on catalogues of parent galaxies of SNe. We have first
obtained from the NASA/IPAC Extragalactic Database (NED) the most
accurate positions for the sources in the Cappellaro et al.'s
database, restricting ourselves to morphological Hubble types $T \leq
-1.5$.  We then cross-correlated this list with the NRAO VLA Sky
Survey (NVSS; Condon et al. 1998), which covers the whole sky north of
$-40^{\circ}$ at 1.4 GHz. For galaxies below this limit we used the
Parkes-MIT-NRAO survey (PMN; Griffith \& Wright 1993) at 4.85 GHz,
converting the flux to 1.4 GHz assuming a radio spectral index equal
to $-0.75$. Sources without an NVSS radio match were dealt with in
two steps. First, to take care of cases where the NVSS might be
resolving out the source or where the core was very weak and most of
the flux was coming from the lobes (which could be way off the optical
position), we cross-correlated our database with the PMN (in the
south) and the GB6 (Gregory et al. 1996; in the north)
catalogues. Second, if still no match was found, we assumed a radio
flux $< 2.5$ mJy, the limit of the NVSS. The adopted correlation
radius was chosen according to the uncertainties on the optical and
radio positions of the sources. Namely, a radius of $20$ and
$30\arcsec$ was taken for the NVSS and PMN/GB6 respectively for
galaxies having optical positional uncertainties $< 5\arcsec$. Radii
of $30$ and $45\arcsec$ were adopted for galaxies having larger
positional errors. Note that this approach is conservative as it might
still miss some interacting galaxies for which the optical position
could be quite different from the radio one.  The NVSS will also
underestimate the total flux for extended radio sources. This cannot
have a major effect on our results for two reasons: 1. we checked all
of the galaxies in Tab. 2 and found only one such case, NGC 1316,
whose classification did not change when using the total flux; 2. we
find a fraction of radio-loud sources $\sim 12\%$ (see below),
perfectly consistent with the value generally quoted (Ledlow \& Owen
1996).  Finally we note that the offset of $\sim 20\arcsec$ between
the optical and the PMN position of E471$-$G27 is consistent with the
PMN positional uncertainty. However, the optical and NVSS positions
differ by $\sim 1.2\arcmin$. Given that NED includes amongst this
source's names PMN J2351$-$2758 and TXS 2349$-$282 and that the NVSS
image shows that the optical source lies at the edge of the radio
contours, we decided to associate a radio source with this galaxy.

Out of a total of 2217 galaxies we have found 267 radio-loud, 212
radio-faint, and 1738 radio-quiet galaxies respectively. The
twenty-one galaxies in which SNe were discovered are reported in
Tab.~2. Col. 1 and 2 give the SN and the galaxy designation, col. 3
the morphological Hubble type, col. 4 the radio flux and col. 5 the
rate of decline of the luminosity expressed as $\Delta m_{15}$ (see
Phillips 1993). Fig. 1 shows the frequency distribution of all 2215
early type galaxies belonging to our sample as a function of their
absolute B-band magnitudes, M$_B$ (solid line). The dashed line refers
to the distribution of the general luminosity function of ellipticals
(adapted from Muriel et. al. 1995); the shaded histogram denotes the
distribution of radio-loud galaxies. The absolute magnitudes of the
galaxies have been taken from the RC3 catalogue (de Vaucouleurs et
al. 1991).  In tab.~1 we report the galaxy type (col. 1), the number
of galaxies for each type used by Della Valle \& Panagia (2003) and in
this paper (col. 2 and col. 5), the normalized control times, \ie the
product of the control time (\ie the time during which a given galaxy
has been observed) to the galaxy's B-band absolute luminosity, so that
is given in units of years $\times$ 10$^{10}$ L$^B_\odot$ (col. 3 and
col. 6), and the number of type Ia SNe discovered (col. 4 and
col. 7). $H_0=75~km~s^{-1}~Mpc^{-1}$ is adopted throughout the paper.

\begin{table}
\begin{center}
\caption{SNe occurred in early type galaxies belonging to the sample }
\label{sample}
\begin{tabular}{lllrr}
\hline
SN  &  Galaxy  & T & radio power   & $\Delta$m$_{15}$\\
    &   NGC    &   &$10^{29}$erg/s Hz$^{-1}$&                 \\
\hline
 1961H	&  4564	 & -4.9&	$<0.008$     &         \\
 1968A  &  1275	 & -1.6&	720   &         \\
 1970J	&  7619	 & -5.0&	0.54  &	 1.30   \\ 
 1972J  &  7634	 & -2.0&	$<0.04$     &         \\
 1980I	&  4374	 & -4.7&	20    &         \\
 1980N	&  1316	 & -1.9&	1000  &  1.28   \\
 1981D	&  1316	 & -1.9&	1000  &  1.33   \\
 1981G	&  4874	 & -4.0&	24    &         \\ 
 1983G	&  4753	 & -2.2&	0.53  &  1.50   \\
 1983J  &  3106	 & -1.9&	0.92  &         \\
 1986G  &  5128	 & -2.2&	18    &  1.73   \\
 1990M  &  5493	 & -2.1&	$<0.04$     &         \\ 
 1991Q  &  4926A & -1.6&	0.31  &         \\
 1991bg &  4374	 & -4.7&	20    &  1.88   \\
 1991bi &  5127	 & -4.9&	8.7   &	        \\
 1992A	&  1380	 & -1.9&	$<0.02$     &         \\
 1992bo	&  E352-G57&-1.5&	$<0.16$     &         \\ 
 1993C  &  2954	 & -4.9&	$<0.08$     &         \\
 1993ah &  E471-G27&-2.0&	41    &  1.30   \\
 1994D  &  4526	 &-2.0& 	0.042 &  1.27   \\ 
 1996X  &  5061	 &-5.0&		$<0.03$     &         \\

\hline
\end{tabular}

\end{center}
\end{table}

An inspection to Fig.1 and Tab.1 reveals the following:

i) The Cappellaro et al. (1999) sample is not biased in favour of bright
galaxies. It shows indeed a deficiency in the number of galaxies
brighter than M$_B=-21$ in comparison with a typical luminosity
function of early type galaxies (e.g.  Muriel, Nicotra \& Lambas
1995).  As apparent from Fig. 1 the distribution of the monitored
galaxies has not the marked deficiency of galaxies fainter than
M$_B$=--19 which characterized the `naked-eye' Evans sample.

ii) The fraction of radio-loud galaxies in our sample is $0.12\pm0.02$
of the whole sample of galaxies.  Such fraction is essentially
identical to the value $0.140\pm0.024$ as reported by Ledlow \& Owen
(1996) for radio galaxies having the same radio luminosities
($>10^{29}$ erg s$^{-1}$ Hz$^{-1}$ at 1.4GHz). This indicates that the
frequency of radio galaxies in our sample is not affected by biases
related to their radio properties.

iii) Considering all early type galaxies, regardless of their radio
properties, an overall SNIa rate of $0.19^{+0.05}_{-0.04}h^2_{75}$
SNU\footnote{1SNU=1 SN(100yr)$^{-1}$ (10$^{10}L^B_\odot$)$^{-1}$} is
obtained.

\section{Analysis and Results}

Since the Cappellaro et al. (1999) sample includes the EVC89, we have
first carried out an independent analysis of the results obtained by
Della Valle \& Panagia, just using C93 (which does not include
EVC89). Separating the galaxies into radio-loud and radio-quiet (this
subclass includes also radio-faint for homogeneity with the previous
paper) we obtain 5.5 and 11.5 SNIa events, respectively. Non-integer
numbers derive from NGC 3106 which is a border line object between the
two classes having a radio luminosity of $\sim 9.2\times 10^{28}$
erg/s/Hz, and therefore we have attributed 0.5 SN to each
bin\footnote{Attributing half SN event to each bin for objects close
to the luminosity threshold may appear a quite unusual procedure,
nevertheless it has been adopted to minimize the effects due to the
uncertainty on H$_\circ$ on the computation of the radio power.  This
approach may affect the measurements of the SN rates in radio -loud
and -quiet galaxies up to $\sim 6\%$ and $10\%$, respectively.} The SN
rate is derived as a ratio between the number of detected SNe and the
respective controltime (expressed in units of 100 years) and it gives:
$r_{loud}=0.41^{+0.25}_{-0.20}h^2_{75}$SNu and
$r_{quiet}=0.17^{+0.07}_{-0.05}h^2_{75}$SNu.  The errors corresponds
to 1$\sigma$ and have been computed from a direct application of
Poisson statistics in the small number regime (e.g. Ebeling 2003,
Gehrels 1986). A {\sl T-Student} test rejects the null hypothesis
between radio-loud and radio-quiet galaxy samples with a probability
of $P\lsim 5\times 10^{-3}$.  This result points out the high
statistical significance of C93 sample itself and provides independent
evidence for the existence of an overproduction of SNe-Ia in
radio-loud galaxies. With this in mind we can now merge the EVC89 and
C93 samples and determine the rate of type Ia SNe in radio-galaxies on
even more robust empirical grounds. The enlarged sample has allowed us
to allocate 3 classes of galaxies on the basis of their radio
activity, as pointed out in Section 1. We have 9.5, 4 and 7.5 SNIa
events in radio-loud, radio-faint and radio-quiet galaxies,
respectively.  Fractional numbers derive from two border lines
objects, NGC 3106 (see above) and NGC 4526 which radio luminosity is
$\sim 4\times 10^{27}$ erg/s/Hz. For the SN rates we obtain:
$r_{loud}=0.43^{+0.19}_{-0.14}h^2_{75}$SNu,
$r_{faint}=0.23^{+0.18}_{-0.11}h^2_{75}$SNu and
$r_{quiet}=0.11^{+0.06}_{-0.03}h^2_{75}$SNu, respectively. The errors
corresponds to 1$\sigma$ and have been computed as above. A {\sl
T-Student} test rejects the null hypothesis between radio-loud and
radio-quiet galaxy samples with a probability of $P\sim 4\times
10^{-4}$.

The comparison among the SN rates, normalized to the B luminosity, in
different Hubble types galaxies should be taken with some degree of
caution because the B luminosity is a poor tracer of the galaxy mass
in stars. In this case, since our analysis is restricted to E/S0
Hubble types (T$\leq -1.5$), this approximation is still acceptable
(see Tab. 3). However, the recent release of the Two Micron All Sky
Survey (2MASS) collaboration (Jarret et al. 2003), allows us to
normalize the rates to the IR luminosity (K-band) and subsequently to
the mass in stars of the parent galaxies by fitting spectrophotometric
models (e.g. Brinchmann \& Ellis 2000; Dickinson et al. 2003) through the
broad band fluxes (see Mannucci et al. 2005 for details). In Tab. 4 we
also report the SN rate per unit of stellar mass.

\begin{table}
\begin{center}
\caption{SN rates in units of SNe per century per $10^{10}$ L$_\odot^B$ }
\label{sample}
\begin{tabular}{lllll}
\hline
              & $rate$ & $1\sigma$        & 2$\sigma$  &3$\sigma$\\         
\hline
radio-loud   & 0.43& $^{+0.19}_{-0.14}$ & $^{+0.38}_{-0.23}$ & $^{+0.60}_{-0.30}$\\
radio-faint  & 0.23& $^{+0.18}_{-0.11}$ & $^{+0.36}_{-0.16}$ & $^{+0.59}_{-0.19}$\\
radio-quiet  & 0.11& $^{+0.06}_{-0.03}$ & $^{+0.11}_{-0.06}$ & $^{+0.18}_{-0.07}$\\
\hline
\end{tabular}
\end{center}
\end{table}

\begin{table}
\begin{center}
\caption{SN rates in units of SNe per century per $10^{10}$ M$_\odot$}
\label{sample}
\begin{tabular}{lllll}
\hline
              & $rate$ & $1\sigma$        & 2$\sigma$  &3$\sigma$\\         
\hline
radio-loud   & 0.100& $^{+0.044}_{-0.032}$ & $^{+0.089}_{-0.054}$ & $^{+0.141}_{-0.070}$\\
radio-faint  & 0.052& $^{+0.041}_{-0.025}$ & $^{+0.082}_{-0.038}$ & $^{+0.135}_{-0.046}$\\
radio-quiet  & 0.023& $^{+0.012}_{-0.008}$ & $^{+0.024}_{-0.013}$ & $^{+0.035}_{-0.018}$\\
all E's      & 0.044& $^{+0.016}_{-0.014}$ &                      &                     \\
S0a/b        & 0.063& $^{+0.027}_{-0.025}$ &                      &                     \\
Sbc/d        & 0.170& $^{+0.068}_{-0.063}$ &                      &                     \\
Irr          & 0.77 & $^{+0.42}_{-0.31}$   &                      &                     \\
\hline
\end{tabular}
\end{center}
\end{table}

\section{Discussion and Conclusions}
\bigskip

Our results allow us to confidently conclude that the SNIa rate in
radio-loud galaxies is a full factor of 4.4$^{+2.4}_{-2.6}$ higher
than it is measured in radio-quiet objects confirming, on more robust
statistical grounds, the findings of Della Valle \& Panagia (2003). We
have also obtained some evidence (P$\lsim 5\times 10^{-2}$) that the
class of radio-faint galaxies may have a rate that is intermediate
between radio-loud and radio-quiet objects. 
This enhancement in SNIa events in radio galaxies can be explained in
basically two different ways:

1) The enhanced SNIa explosion rate in radio-loud galaxies is a {\it
direct} consequence of their being strong radio sources. This is the
scenario suggested by Capetti (2002) and Livio et
al. (2002). According to these authors the accretion rate from the
interstellar medium onto white dwarfs could be enhanced by the action
of the jets up to either driving an accreting white dwarf to approach
the Chandrasekhar mass and trigger a SNIa explosion (Livio 2000, and
references therein) or increasing the accretion rate from the
interstellar medium onto the white dwarfs by 1-2 orders of magnitude,
up to $\sim 10^{-8/-9}$M$_\odot$ yr$^{-1}$, which is the rate
necessary to trigger classical nova explosions (Prialnik \& Kovetz
1995).  This process might account for the high rate of novae observed
in M87 in the regions nearby the jets. However, we note that equation
(4) of Livio et al (2002), which is the transcription of Bondi's
(1952) formula for accretion on a star at rest, is valid only for very
low stellar velocities ($<<1~$km~s$^{-1}$). This fact has important
consequences on the practicability of this scenario. In a rather
optimistic scenario in which $v_{star}=1~km~s^{-1}$ and the jet has a
width of $\sim$100~pc, the transit time is about 100~Myrs during which
the total accreted mass would be less than 0.1M$_\odot$, i.e.  hardly
enough to make a WD explode as SNIa, unless its mass was greater than
$\sim 1.35M_\odot$.  In a more realistic case, in which one assumes
typical star velocities of the order of 10-100 km~s$^{-1}$ (see for
example Williams 1994, who has measured the velocities of nova systems
in the Milky Way) the accreted mass onto massive WDs would be only
$10^{-5}$--$10^{-8}$ $M_\odot$. This makes the accretion process from
ISM highly inefficient to trigger SNIa explosions (while it may play a
not negligible role to trigger nova explosions) even in the presence
of a high compression regime. In addition, we note that: \\ {\sl i)}
Even under the best conditions that could induce a SNIa explosion
($v<<1$~km~s$^{-1}$; M$_{WD}\sim 1.2\div 1.3$ M$_\odot$), the
resulting SNIa is expected to be spatially confined to the regions
immediately adjacent to radio jets and/or the bulk of radio activity.
While this may be true for NGC 4374 (Fig. 2), it appears not to be the
case for Fornax~A=NGC~1316 (Fig. 3) in which the two SNIa (SN 1980N
and SN 1981D) are located quite far from the strong radio lobes
(Geldzahler \& Fomalont 1984), nor for Centaurus~A=NGC~5128, in which
SN~1986G is deeply embedded in the equatorial lane of gas and dust
(Fig. 4). \\ {\sl ii)} If the dominant phenomenon is accretion from
the ISM rather than from a binary companion, an extreme consequence is
that all {\it single} WDs could contribute to the Nova and SNIa
production, and not only some binary systems with suitable parameters.
Actually, since the fraction of stars that may give rise to a SNIa is
only $\eta \sim 5-10\%$ of the total number of stars with masses above
3 \msun~(Madau, Della Valle \& Panagia~1998, Dahlen et al. 2004), in
order to explain the observed enhancement in the SN rate (a factor
$\sim 4$) one can work out that jets should be capable of raising the
efficiency of the SN explosions up to $\sim 100\%$ not only in the
regions near the jets, but also over a significant fraction of the
entire galaxy, according to the simple relation:
$\eta_{jet}/\eta_{gal} \sim 3\times vol_{gal}/vol_{jet}$ where $\eta$
parameter is the efficiency of SN explosions and $vol_{jet}$ and
$vol_{gal}$ are the volumes sampled by the jet and the whole
galaxy. \\ {\sl iii)} For a density of 100~cm$^{-3}$ and a jet width
of 100~pc as advocated by Livio et al. (2002), the column density of
HI associated with a jet would be about $3\times
10^{22}$~cm$^{-2}$. In this case one should expect an intense 21cm
line emission, which has never been observed in association with
jets. \\ {\sl iv)} A clear indication that accretion of interstellar
matter by slow (v$<<1~km~s^{-1}$) WDs is not an efficient process to
trigger SNIa explosions, is that SNIa occurring in spiral galaxies are
not associated with molecular clouds, which are regions with the
``ideal" conditions, i.e. high particle density and very low
temperatures, for triggering a SNIa explosion through this
mechanism. It appears that the accretion from ISM can work only in the
presence of very special circumstances, which makes a SNIa explosion
through accretion ``\`a la Bondi" an exception rather than the rule.
\\ In summary, since there is no evidence that direct accretion from
the ISM could be capable to boost the SNIa rate in radiogalaxies by a
factor of 4, we conclude that the dominant process is very likely a
different one.

2) As an alternative we propose that the enhancement of SNIa explosion
rate in radio-loud galaxies has the {\it same common} origin as their
being strong radio sources (see also Kochhar 1989; Kochhar \& Prabhu
1984), but that there is no {\sl causality} link between the two
phenomena. If the radio activity of a galaxy is triggered by
interaction and/or mergers (e.g. Baade \& Minkowski 1954, Balick \&
Heckman 1982, Heckman \etal~1986), then the SNIa rate can be enhanced
by the formation or the capture of relatively young stellar
populations in which SNIa occur at much higher rates than in genuinely
old populations (Della Valle \& Livio 1994, Panagia 2000, Mannucci et
al. 2005). SNIa progenitors are believed to be stars with original
masses $\gsim$3~\msun~ and $\lsim$8~\msun (Greggio \& Renzini 1983)
having nuclear-burning lifetimes shorter than $\sim$400~Myrs. After
that time, the star becomes a white dwarf and may explode as a SNIa
only after an additional time as needed for either accreting mass from
a companion or merging with it. Such a time ($\tau$) is essentially
unconstrained by theory, but measurements of SNIa rates in the local
universe (Mannucci et al. 2005) have shown that: (a) the rate of type
Ia SNe in blue galaxies is about a factor 30 higher than it has been
measured in red galaxies; (b) the rate of type Ia SNe can be
reproduced by combining a constant contribution (corresponding to the
rate of Ia in radio-quiet galaxies) + another contribution which
exhibits a behaviour similar to that of core-collapse SNe (see their
Fig. 6). This implies evolutionary times (main sequence
life-time + $\tau$ $\lsim 0.2$ Gyrs) much shorter than the time scale
needed by young `blue' galaxies to become `red' galaxies ($\sim 0.8$
Gyrs to change (B-K) by about 1 mag, see for example Bruzual \&
Charlot 2003). This result may not be at variance with the recent
findings of Gal-Yam \& Maoz (2004) and Strogler et al.(2004), who
derived $\tau \sim 2-4$ Gyrs from counts of SNeI-a at high z.  This
may rather suggest that $\tau$ spans a broad range of values, possibly
from $\lsim 0.2$ up to a few Gyrs (Mannucci, Della Valle \& Panagia,
in preparation, Greggio 2005).  As a consequence, in order to sustain
a relatively high rate of SNIa explosions (characterized by a short
$\tau$) during the time in which a galaxy is radio-loud ($\Delta
T_{RL}\sim 10^8$ yr, e.g. Srianand \& Gopal-Krishna 1998), a galaxy
has to have a new injection of young stellar populations at suitable
time intervals. Since the fraction of radio-loud ellipticals is about
14\% of the total number of ellipticals (Ledlow \& Owen 1996), we can
estimate that, on average, an elliptical galaxy undergoes a number of
star formation episodes, $n_{inj}$, of the order of $n_{inj}\simeq0.14\times
T_H/\Delta_{RL}\sim 20$, for a Hubble~time $T_H\simeq 14$~Gyrs.
Therefore, the time interval between successive episodes of injection
of young stellar populations, $\Delta t_{inj}$, turns out to be
$\Delta t_{inj}=T_H/n_{inj}\simeq$ 0.7~Gyr. In late type galaxies, the
ongoing star formation provides the needed input for a steady
production of SNIa events. Such a supply in early type galaxies can
naturally be provided by repeated episodes of interaction or mergers
that induces either formation of young stellar populations (galaxy
interaction; \eg the Antennae galaxies, Whitmore \etal~1999) or
capture of young stars from dwarf companions.  Well known examples of
interacting and/or merging early type galaxies are
Centaurus~A=NGC~5128 and Fornax~A=NGC~1316, which are strong radio
sources and indeed have produced 1+2 SNIa in the last century.

Strong support to this hypothesis comes from the following two facts: 

i) In Tab. 4 we have reported the SN rate for Ellipticals radio- loud,
faint, quiet, and Spirals normalized per unit of mass. A simple glance
at col. 2 and Fig. 5 shows that the rates of SNIa are higher in late
type galaxies than in early type galaxies by a full order of
magnitude, then suggesting a connection between the production of type
Ia SNe and star formation activity (e.g. Mannucci et al. 2005). Fig. 5
shows the behaviour of the normalized SNIa rate (per 10$^{10}M_\odot$)
as a function of the morphological Hubble type and radio activity. The
data clearly shows that for the SNIa production, radio-loud early type
galaxies behave like `S0a/b' which are systems where the star forming
activity is certantly dimmed with respect to Sd/Irr, but still active,
likely of the order of $0.1M_\odot$ yr$^{-1}$ per $10^{10}M_\odot$
(e.g. Tinsley 1974, Sandage 1986, Kennicutt 1998). It is then
plausible that this enhancement of SNIa rate originates from episodes
of star formation, which may have been triggered, on time-scale of 1
Gyr, by merging dwarf galaxies. This scenario provides a natural
explanation of why among early type galaxies SNIa rates are higher in
radiogalaxies without requiring the concurrence of any new and/or
additional process to enhance the SNIa rates.

ii) The existence of systematic differences between type Ia SNe
occuring in Spirals and Ellipticals, which has been accumulating in
the last decade both spectroscopically (Branch \& van den Bergh 1993,
Li et al. 2001) and photometrically (van den Bergh \& Pazder 1992,
Della Valle \& Panagia 1992, Hamuy et al. 1995, Hamuy et al. 2000),
can be effectively represented by the distribution of the decline
rate parameter $\Delta m_{15}$ (Phillips 1993; Phillips et
al. 1999). In Fig. 6 we report the distribution of the rates of
decline for late (shade-dotted histogram) and early type
galaxies. Data were taken from Branch, Romanishin and Baron (1996) and
Parodi et al. (2000).  The probability that the two distributions
originate from the same population, is excluded by KS-test at the
level of P=$5\times 10^{-7}$ (see also Altavilla et al. 2004). The
most straightforward interpretation of the double-peaked distribution
of the rates of decline is obtained in terms of an age sequence,
(e.g. Branch \& van den Bergh 1993; Ruiz-Lapuente 1995; Hamuy et
al.1996; Howell 2001; but see the case of SN 2002bo, Benetti et
al. 2003), with metallicity being excluded or at least less important
(Ivanov, Hamuy and Pinto 2000). With the dashed-solid box we have
indicated the rates of decline for the 8 objects belonging to our
sample of SNe in radio-galaxies for which it was possible to derive
the rates of decline. We find tantalizing evidence (P$\lsim 5\times
10^{-2}$) that the rates of decline for the objects occurred in
radio-galaxies are `intermediate' between those of the late and early
type galaxies. This fact supports the existence of a {\sl continuum}
of properties between galaxies with very active (Irr and late
Spirals), active (early Spirals and radio-active Ellipticals) and
exhausted (Ellipticals radio-quiet) star formation rate and type Ia
supernovae. In Spirals with high star forming activity SNIa are
generally overluminous, believed to be produced in younger systems
from the most massive SNIa progenitors (\eg Tutukov \& Yungelson 1994,
1996, Ruiz-Lapuente, Burkert \& Canal 1995), SNeIa in radio-loud
ellipticals are characterized by faster rates of decline (\ie less
luminous SNe) and they might originate from progenitor of intermediate
mass.  Finally radio-quiet ellipticals mainly produce faint type Ia
events from a genuine `old' Pop II stellar population (light
progenitors). An important consequence of our scenario is the
predicted occurrence of a few type II-Ib/c SNe in early type galaxies
at the early stages of the same star formation burst that provides a
steady supply of SNIa progenitors.  Since core collapse SNe are
produced by progenitors more massive than 8M$_\odot$, whose lifetime
is shorter than about 30 Myrs, only for a small fraction of the time
between subsequent interaction/capture episodes (approximately
30Myrs/1Gyr=3\%) would one expect to see such SNe in early type
galaxies.  Since in galaxies with active star formation the rate of
type II and Ib/c SNe is about 2 and 2/3 times than that of SNIa,
respectively, we estimate that one should detect core-collapse ({\sl
CC}) events at a rate 3\%$\times 8/3\simeq$ 8\% of SNIa events, which
corresponds to an average rate of {\sl CC} SNe of about
0.0152$h^2_{75}$SNu.  Given the control time measured by Cappellaro et
al. (1999) for type {\sl II plateau, II linear} and {\sl Ib/c}: 4021
yr, 3213 yr and 5011 yr $\times L^B/10^{10}L_\odot^B$, and the
relative frequencies of occurence of each {\sl CC} SN into the {\sl
plateau}, {\sl linear} and {\sl Ib/c} sub-types (0.6, 0.2 and 0.2
respectively), in our sample one should have detected $0.0152\times$
(4057)/100=0.62 SNe, where we have used the weighted mean of the
control times reported above. Applying Poisson statistics, we estimate
that the probability of finding no {\sl CC} SN event in our sample is
$54\%$, so that the absence of {\sl CC} SNe is just a rather natural
occurrence. We conclude that our prediction will be observationally
testable when the SNIa statistics will have increased the control time
by a factor $\sim 6$ ($\sim 2\sigma$ confidence level).
\smallskip

Our main conclusion is that the type Ia Supernova rate for radio-loud
galaxies is more than a factor of four higher than for radio-quiet
ones. This is true even when the rate is normalized to stellar mass
instead of B-band luminosity, a more physical approach. Combined with
the well-established connection between type Ia SNe and star-formation
rate, this suggests a link between the radio-loud phenomenon and
increased star-formation, which might prove to be an important piece
of the still unexplained puzzle of radio-loudness in the Universe.

\section{Acknowledgments}

MDV is grateful to the Space Telescope Science Institute for the
friendly hospitality and creative atmosphere. This research has made
use of data from the NASA/IPAC Extragalactic Database (NED), which is
operated by the Jet Propulsion Laboratory, California Institute of
Technology, under contract with the National Aeronautics and Space
Administration. The authors wish to thank the referee Alessandro Capetti
for his valuable comments, which have helped to improve the
presentation of this paper.

\newpage
\section{Captions}

{\bf Fig. 1} Distribution of early-type galaxies in the Cappellaro et
al.  (1999) sample as a function of their absolute B-band magnitudes.
The solid line is the distribution of the entire sample, the shaded
histogram is the distribution of radio-loud galaxies, and the dashed
histogram is the general luminosity function of elliptical galaxies,
adapted from Muriel et al.(1995). 
\bigskip

{\bf Fig. 2} Location of SN 1991bg and SN 1957B in NGC 4374. The Image
is a collage of pictures of the galaxy and jets obtained from
Digitalized Sky Survey (DDS) and NASA/IAPC Extragalactic Database
(NED), respectively.
\bigskip

{\bf Fig. 3} Location of SN 1980N and SN 1981D in NGC 1316. Collage of
images obtained from Digitalized Sky Survey (DDS) and NASA/IAPC
Extragalactic Database (NED), respectively.
\bigskip

{\bf Fig. 4} Location of SN 1986G in NGC 5128.  Collage of images
obtained from Digitalized Sky Survey (DDS) and NASA/IAPC Extragalactic
Database (NED), respectively.
\bigskip

{\bf Fig. 5} The rates  of SNIa explosions (normalized to 10$^{10}$
M$_\odot$) as a function of the morphological Hubble type of the
respective parent galaxies. Open symbols at T$\sim -2$ represent the SNIa
rates of radio-loud (top) and radio-quiet (bottom) early-type galaxies.
The filled symbol is the overall rate of SNIa events in early type
galaxies, regardless of their radio properties (i.e. average of radio
-loud and -quiet rates) 
\bigskip

{\bf Fig. 6} The distribution of the light curve rates of decline for
late (shade-dotted histogram) and early type galaxies (solid histogram)
and the respective cumulative distributions.
The dashed box  indicates the rates of decline for 8 objects belonging
to the sample listed in Tab. 2.


\vskip .3in

\begin{thebibliography}{}
\bibitem[]{} Altavilla, G., Fiorentino, G., Marconi, M. et al. 2004, MNRAS, 349, 1344
\bibitem[]{} Baade, W., \& Minkowski, R 1954, ApJ, 119, 206
\bibitem[]{} Balick, B., Heckman, T.M., 1982, ARA\&A, 20, 431
\bibitem[]{} Benetti, S., Meikle, P., Stehle, M. et al. 2004, MNRAS, 348, 261 
\bibitem[]{} Bondi, H. 1952, MNRAS, 112, 195
\bibitem[]{} Branch, D., \& van den Bergh, S. 1993,  AJ, 105, 2231
\bibitem[]{} Branch, D., Romanishin, W., \& Baron, E. 1996, 470, L7
\bibitem[]{} Bruzual, G. \& Charlot, S. 2003, MNRAS, 344, 1000
\bibitem[]{} Brinchmann, J., \& Ellis, R.S. 2000 ApJ, 536, L77
\bibitem[]{} Capetti, A., 2002, ApJ, 574, L25
\bibitem[]{} Cappellaro, E., Turatto, M., Benetti, S., Tsvetkov, D. Yu., Bartunov, O. S., 
\& Makarova, I. N 1993, A\&A, 268, 472
\bibitem[]{} Cappellaro, E., Evans, R., \&  Turatto, M. 1999, A\&A, 351, 459
\bibitem[]{} Condon, J. J., Cotton, W. D., Greisen, E. W., Yin, Q. F., Perley, R. A., 
Taylor, G. B., \& Broderick, J. J. 1998, AJ, 115, 1693
\bibitem[]{} Dahlen et al. 2004, ApJ, 613, 189 
\bibitem[]{} de Vaucouleurs G., de Vaucouleurs A., Corwin H.G., et al., 1991, 
Third Reference Catalogue of Bright Galaxies. Springer-Verlag, New York [RC3]
\bibitem[]{} Della Valle, M., \& Livio, M., 1994, ApJ,423, L31
\bibitem[]{} Della Valle, M., \& Panagia, N., 1992, AJ, 104, 696
\bibitem[]{} Della Valle, M. \& Panagia, N. 2003, ApJ, 587, L71 
\bibitem[]{} Dickinson, M., Papovich, C., Ferguson, H. C., \& Budavari, T. 2003, 
ApJ, 587, 25
\bibitem[]{} Ebeling, H. 2003, MNRAS, 340, 1269
\bibitem[]{} Evans, R, van den Bergh, S., \& Clure, R.D., 1989, ApJ, 345, 752
\bibitem[]{} Gal-Yam, A., \& Maoz, D. 2004, MNRAS, 347, 942
\bibitem[]{} Gehrels, N. 1986, ApJ, 303, 336
\bibitem[]{} Geldzahler, B. J., \& Fomalont, E. B., 1984, AJ, 89, 1650
\bibitem[]{} Grggio, L. 2005, A\&A, submitted
\bibitem[]{} Gregory, P.C., Scott, W.K., Douglas, K., \& Condon, J.J.
1996, ApJS, 103,427
\bibitem[]{} Griffith, M. R., \& Wright, A. E. 1993, AJ, 105, 1666
\bibitem[]{} Greggio, L., \& Renzini, A. 1983, A\&A, 118, 217
\bibitem[]{} Hamuy, M., Phillips, M.M., Maza, J., Suntzeff, N.B.,
Schommer, R.A.,  \& Aviles, R. 1995, AJ, 109,1
\bibitem[]{} Hamuy, M., Phillips, M. M., Suntzeff, N.B., Schommer, R.
A.; Maza, J., \& Aviles, R. 1996, AJ, 112, 2391
\bibitem[]{} Hamuy, M., Trager, S. C., Pinto, P.A., Phillips, M. M.,
Schommer, R. A., Ivanov, V., \& Suntzeff, N.B. 2000, AJ, 120, 1479
\bibitem[]{} Hamuy, M., Phillips, M. M., Suntzeff, N. B. et al. 2003
Nature, 424, 651
\bibitem[]{} Heckman, T.M., Smith, E.P., Baum, S.A., van Breugel,
W.J.M., Miley, G.K., Illingworth, G.D., Bothun, G.D., \& Balick, B.
1986, ApJ, 311, 526
\bibitem[]{} Howell, D.A., 2001, ApJ, 554, L193
\bibitem[]{} Ivanov, V.D., Hamuy, M., \& Pinto, P.A. 2000, ApJ, 542, 588
\bibitem[]{} Jarrett, T. H.; Chester, T.; Cutri, R.; Schneider, S. E.;
\& Huchra, J. P 2003, AJ, 125, 525
\bibitem[]{} Kennicutt, R.C., ARA\&A, 36, 189
\bibitem[]{} Kochhar, R.K. 1989, Astrophys. and Space Sci., 157, 305 
\bibitem[]{} Kochhar, R.K., \& Prabhu, T.P. 1984, Astrophys. and Space
Sci., 100, 369
\bibitem[]{} Ledlow, M.J., \& Owen, F.N., 1996, AJ, 112, 9
\bibitem[]{} Li, W., Filippenko, A. V., Treffers, R. R.; Riess, A. G.,
Hu, J., \& Qiu, Y. 2001, ApJ, 546, 734
\bibitem[]{} Livio, M., 2000, in Type Ia Supernovae: Theory and
Cosmology, eds. J.C. Nyemeyer, \& J.W. Truran (Cambridge: Cambridge
Univ. Press), p. 33
\bibitem[]{} Livio, M., Riess, A., \& Sparks, W., 2002, ApJ, 571, L99
\bibitem[]{} Livio, M., \& Riess, A. 2003, ApJ, 594, L93
\bibitem[]{} Madau, P., Della Valle, M., \& Panagia, N., 1998, MNRAS, 297, L17
\bibitem[]{} Mannucci, F., Della Valle, M., Panagia, N., Cappellaro, E.,
Cresci, G.,  Maiolino, R., Petrosian, A., \& Turatto, M. 2005, A\&A, in
press (astro-ph/0411450)
\bibitem[]{} Maoz, D.,  \& Gal-Yam, A. 2004, MNRAS, 347, 951
\bibitem[]{} Muriel, H. Nicotra, M. A., \& Lambas, D.G. 1995, AJ, 110, 1032
\bibitem[]{} Panagia, N., 2000, in Experimental Physics of Gravitational
Waves, eds. M. Barone, G. Calamai, M. Mazzoni, R. Stanga and F. Vetrano,
(Singapore: World Scientific Publishing Co.), p. 107.
\bibitem[]{} Parodi, B.R., Saha, A., Sandage, A., \& Tammann, G. A. 2000,
ApJ, 540, 634
\bibitem[]{} Phillips, M. M., Jenkins, C. R., Dopita, M. A., Sadler, E.
M., \& Binette, L. 1986,  AJ, 91, 1062
\bibitem[]{} Phillips, M.M. 1993, ApJ, 413, L105
\bibitem[]{} Phillips, M.M., Lira, P., Suntzeff, N.B., Schommer, R.A.,
Hamuy, M., \& Maza, J. 1999, AJ, 118, 1766
\bibitem[]{} Prialnik, T., \& Kovetz,A. 1995, ApJ, 445, 789 
\bibitem[]{} Renzini, A., Ciotti, L., D'Ercole, A., \& Pellegrini, S.
1993, ApJ, 419, 52
\bibitem[]{} Ruiz-Lapuente, P., Burkert, A., \& Canal, R. 1995, ApJ, 447, L69
\bibitem[]{} Sadler, E.M., Jenkins, C.R., \& Kotanyi C.G. 1989, MNRAS, 240, 591
\bibitem[]{} Sandage, A. 1986, A\&A, 161, 89
\bibitem[]{} Srianand, R., \& Gopal-Krishna 1998, A\&A, 334, 39
\bibitem[]{} Strolger et al. 2004, ApJ, 613, 200
\bibitem[]{} Tinsley, B.M. 1974, A\&A, 31, 463
\bibitem[]{} Tutukov, A.V., \& Yungelson, L. 1994, MNRAS, 268, 871
\bibitem[]{} Tutukov, A.V., \& Yungelson, L. 1996, MNRAS, 280, 1035
\bibitem[]{} Urry, C.M., \& Padovani, P. 1995, PASP, 107, 803   
\bibitem[]{} van den Bergh, S.,\& Padzer J., ApJ, 390, 34
\bibitem[]{} Whitmore, B.C., Zhang, Q. Leitherer, C., Fall, S. M.,
Schweizer, F., \& Miller, B.W. 1999, AJ, 118, 1551
\bibitem[]{} Williams, R.E. 1994, ApJ, 426, 279
\end{thebibliography}
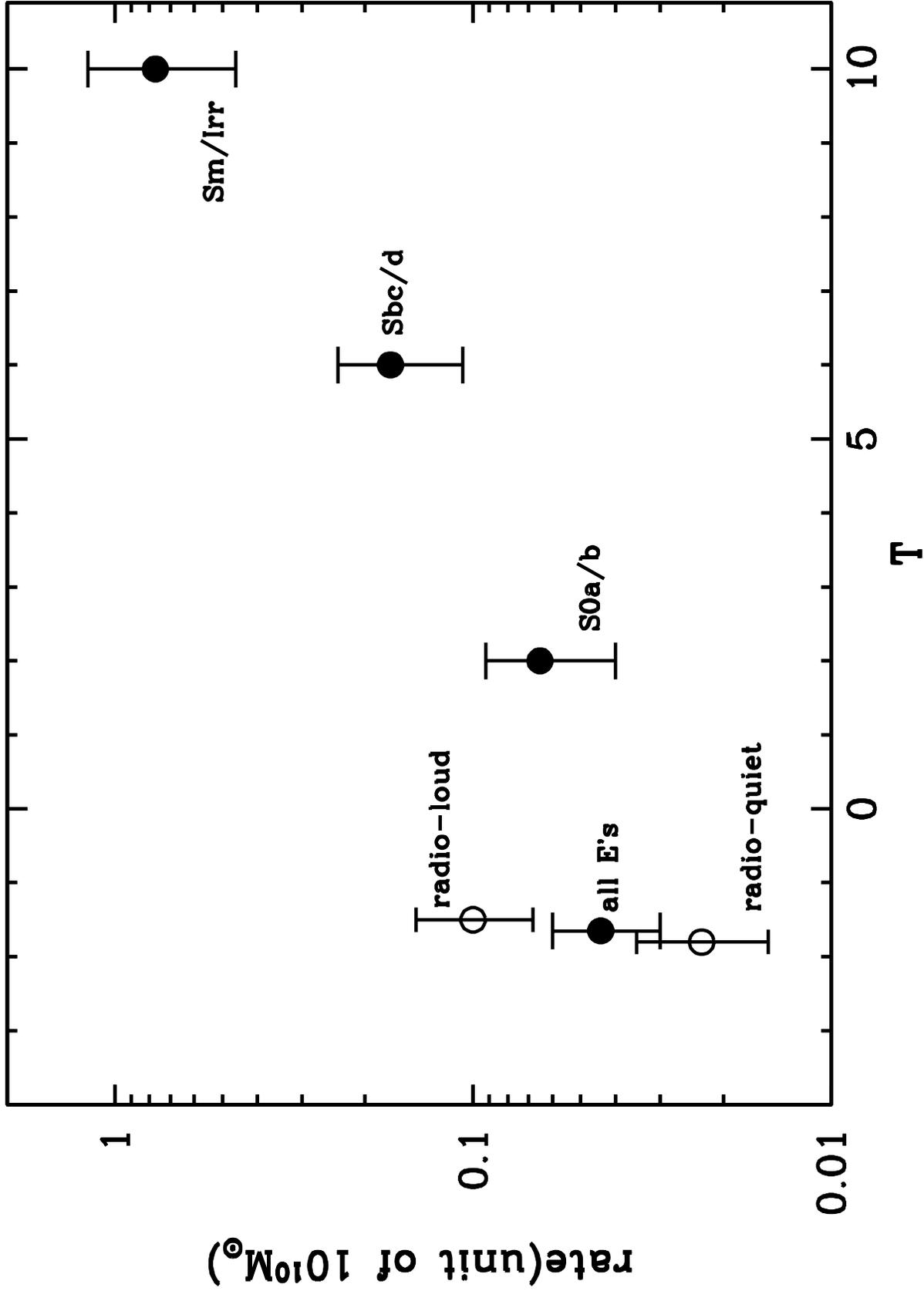
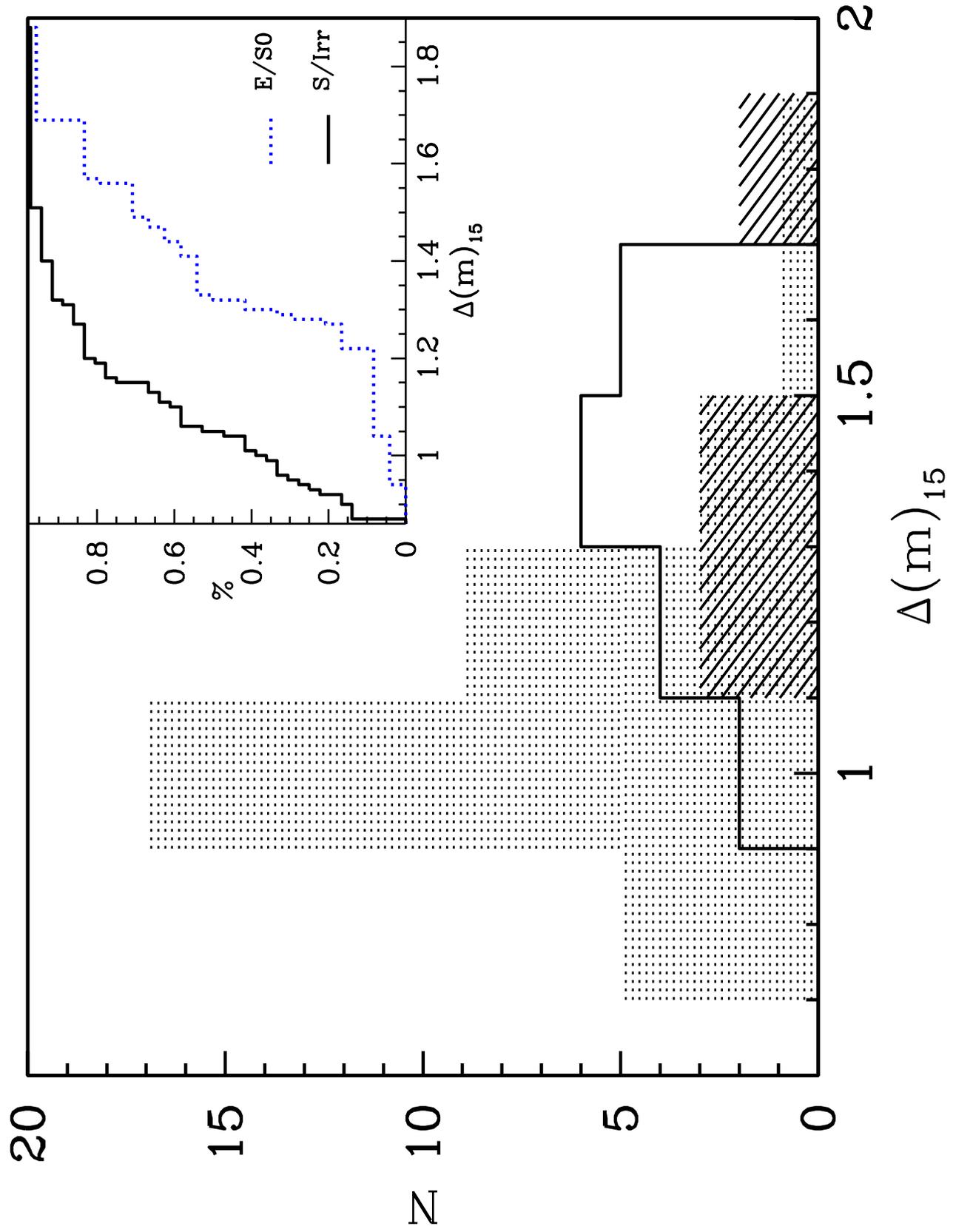
\end{document}